\documentclass[journal]{IEEEtran}
\usepackage{cite}
\usepackage{amsmath,amssymb,amsfonts}
\usepackage{graphicx}
\usepackage{textcomp}
\usepackage{soul}
\usepackage{color}
\usepackage{booktabs}
\usepackage{algorithm}
\usepackage{algorithmic}

\usepackage[table,dvipsnames]{xcolor}
\usepackage[colorlinks,
            linkcolor=blue,
            anchorcolor=blue,
            citecolor=blue,
            urlcolor=blue
            ]{hyperref}
\usepackage{multirow}

\usepackage{xcolor}
\def\BibTeX{{\rm B\kern-.05em{\sc i\kern-.025em b}\kern-.08em
    T\kern-.1667em\lower.7ex\hbox{E}\kern-.125emX}}

\newcommand{\blk}{\color{black}}

\begin{document}
\title{Learning-Based Design of LQG Controllers in Quantum Coherent Feedback}
\author{Chunxiang~Song,
        Yanan~Liu,
        Guofeng~Zhang,
        Huadong~Mo,
        and~Daoyi~Dong
\thanks{This work has been submitted to the IEEE for possible publication. Copyright may be transferred without notice, after which this version may no longer be accessible.}
\thanks{C. Song is with the School of Engineering and Technology, University of New South Wales, Canberra, ACT 2600, Australia
        {(\tt\small chunxsong@gmail.com).}}
\thanks{Y. Liu is with the School of Engineering, University of Newcastle, Newcastle, NSW 2308, Australia, and with Institute for Automatic Control and Complex Systems (AKS), University of Duisburg-Essen, 47057 Duisburg, Germany {(\tt\small yaananliu@gmail.com).}}
\thanks{G. Zhang is with the Department of Applied Mathematics, The Hong Kong Polytechnic University, Hung Hom, Kowloon, Hong Kong Special Administrative Region of
China, and with Shenzhen Research Institute, The Hong Kong Polytechnic University, Shenzhen 518057, China {(\tt\small guofeng.zhang@polyu.edu.hk).}}
\thanks{Huadong Mo is with the School of Systems and Computing, University of New South Wales, Canberra, ACT 2600, Australia {(\tt\small huadong.mo@unsw.edu.au).}}
\thanks{D. Dong is with the Australian Artificial Intelligence Institute, Faculty of Engineering and Information Technology, University of Technology Sydney, Sydney, NSW 2007, Australia {(\tt\small daoyidong@gmail.com).}}
}

\maketitle

\begin{abstract}
In this paper, we propose a differential evolution (DE) algorithm specifically tailored for the design of Linear-Quadratic-Gaussian (LQG) controllers in quantum systems. Building upon the foundational DE framework, the algorithm incorporates specialized modules, including relaxed feasibility rules, a scheduled penalty function, adaptive search range adjustment, and the ``bet-and-run'' initialization strategy. These enhancements improve the algorithm's exploration and exploitation capabilities while addressing the unique physical realizability requirements of quantum systems. The proposed method is applied to a quantum optical system, where three distinct controllers with varying configurations relative to the plant are designed. The resulting controllers demonstrate superior performance, achieving lower LQG performance indices compared to existing approaches. Additionally, the algorithm ensures that the designs comply with physical realizability constraints, guaranteeing compatibility with practical quantum platforms. The proposed approach holds significant potential for application to other linear quantum systems in performance optimization tasks subject to physically feasible constraints.
\end{abstract}


\section{Introduction}
\label{sec:introduction}
\IEEEPARstart{Q}{uantum} optical systems are becoming foundational in advancing quantum technologies, serving as essential components for quantum computing, communication, and precision measurement \cite{gerry2023introductory}. These systems aim to surpass the capabilities of classical technologies, particularly through applications modeled by quantum stochastic differential equations (QSDEs), which provide an effective approximation of underlying field dynamics in quantum optics \cite{james2008h,dong2010quantum,liu2021fault}. Unlike classical systems, quantum systems are subject to stringent physical realizability constraints, as established by James et al. \cite{james2008h}, which ensure that system matrices accurately represent physical quantum behaviors.

Quantum feedback control is crucial in quantum technologies, focusing on the dynamic interactions between a quantum plant and a controller to achieve desired system performance \cite{dong2022quantum}. Controllers may be quantum, classical, or hybrid. Classical controllers typically employ measurement-based feedback, which is widely applied in areas such as quantum error correction \cite{ahn2002continuous}, state purification \cite{combes2008rapid}, and target state stabilization \cite{liu2016lyapunov, kuang2021rapid,song2024measurement,zhou2025auxiliary}. However, measurements inherently introduce coherence loss, affecting the integrity of the quantum state.
In contrast, when the controller itself is a fully quantum system, it enables coherent feedback control. Here, quantum information is exchanged with the plant directly through shared electromagnetic fields, lasers, or even direct coupling, without the need for measurement-based processes. This “coherent” approach preserves quantum coherence, showing promise in achieving superior performance metrics and faster response times compared to classical feedback systems \cite{hamerly2012advantages,jacobs2014coherent}.

The quantum coherent linear quadratic Gaussian (LQG) control problem has been extensively studied in recent years. Previous research \cite{nurdin2009coherent} has shown that a fully quantum linear controller, designed to satisfy physical realizability conditions, can outperform classical LQG controllers in closed-loop configurations. Based on this work, various controller design approaches have been proposed to address quantum coherent LQG control problems, including a Newton-like optimization scheme \cite{vladimirov2013quasi} and a gradient descent-based numerical method \cite{sichani2017numerical} for optimal controller synthesis. \blk To further enhance control performance, Zhang et al. \cite{zhang2010direct} introduced a direct coupling strategy, which facilitates direct energy exchange between the plant and the controller, thereby improving the controller’s effectiveness. Building on this, Zhang et al. \cite{zhang2012coherent} employed signal squeezing techniques on input signals, achieving additional performance gains. Despite these advances, designing quantum controllers for high-dimensional quantum systems remains challenging. This task often involves complex rank-constrained linear matrix inequalities (LMI) with non-convex, nonlinear constraints, making the optimization process highly intricate.

With the growing role of artificial intelligence, evolutionary algorithms (EAs) are increasingly valued for their capability to handle complex optimization problems using population-based strategies. Inspired by natural selection, EAs are effective in solving nonlinear, multi-objective, and high-dimensional constraint problems that defy traditional methods. Their global optimization capability and robustness in high-dimensional search spaces have led to applications in quantum control. In particular, EAs play a key role in quantum learning control, where they are used to optimize control policies and identify robust solutions under uncertainty \cite{brif2010control,judson1992teaching,dong2023learning,dong2019learning,dong2015sampling,sharma2010genetic}. For example, genetic algorithm (GA) has been used to design laser pulses for molecular state transitions \cite{sharma2010genetic}, and differential evolution (DE) has been applied to identify robust regions in quantum control problems \cite{dong2019learning}.

To address non-convex and nonlinear constraints inherent in quantum coherent controllers, \cite{harno2014synthesis} introduced a DE-based optimization method to construct optimal linear coherent quantum controllers. This approach is validated in a quantum network entanglement control example involving two cascaded optical parametric amplifiers. Building on this foundation, we develop an enhanced DE-based framework with multiple improvements tailored for quantum coherent control, enabling efficient optimization across different controller design configurations.

The main contributions of this work are summarized as follows:  
\begin{enumerate}{}{} 
    \item A tailored DE algorithm is developed, integrating {relaxed feasibility rules}, {scheduled dynamic penalty}, {adaptive range adjustment}, and a {``bet-and-run'' initialization} to handle non-convex constraints in quantum coherent control.  
    
    \item A unified framework for fully quantum LQG controllers is established, addressing Lyapunov stability and quantum commutation constraints. The framework extends to three different controller configurations: 
    \begin{itemize}
        \item Indirect coupling via shared quantum fields,
        \item Direct coupling enabling coherent energy exchange,
        \item Hybrid coupling with ideal squeezers for noise suppression.
    \end{itemize}

    \item Numerical experiments demonstrate improved performance indexes and constraint satisfaction across controller configurations, while ablation studies confirm the necessity of key algorithmic components. 
\end{enumerate}  

The structure of this paper is as follows. Section~\ref{problemformulation} provides a brief introduction to the non-commutative stochastic dynamic model for linear quantum systems and formulates the control problem as a constrained optimization task. Section~\ref{sec:QCTDE} details the proposed algorithm based on the DE approach to address such problems. Section~\ref{sec:Apps} presents the implementation of the proposed algorithm in various quantum coherent control scenarios, along with numerical results and an ablation study to analyze its functionality and performance. Finally, Section~\ref{sec:conclusion} concludes the paper.

\textit{Notation}:
Let $i = \sqrt{-1}$ be the imaginary unit. $\mathbb{I}$ is the identity matrix. The commutator is defined by $[A, B] = AB-BA$. The symbol ``diag'' is a block diagonal matrix, assembled from the given entries. The notation $\langle \bullet \rangle$ refers to quantum expectation.

\section{System Model and Problem Formulation}\label{problemformulation}
\subsection{System Model}
\noindent In this section, we consider a general class of linear quantum systems, whose dynamics can be described by the following non-commutative stochastic dynamic model:
\begin{equation}\label{eq:generalSS}
\begin{aligned}
dx(t) &= Ax(t)\,dt + B\,dw(t), \quad x(0) = x_0, \\
dy(t) &= Cx(t)\,dt + D\,dw(t).
\end{aligned}
\end{equation}
Here the system variables, $x(t) = [x_1(t), \dots, x_n(t)]^T$, are a vector of self-adjoint, possibly non-commutative operators. For example, it can be a vector of amplitude and phase quadrature operators for different modes in a continuous-wave (CW) quantum system. In this paper, we only consider the case that $A\in \mathbb{R}^{n\times n}, B\in\mathbb{R}^{n\times n_w}, C\in\mathbb{R}^{n_y\times n}$, and $D\in \mathbb{R}^{n_y\times n_w}$ are real matrices. For the quantum systems described by a vector of annihilation and creation operators, where the corresponding matrices are normally complex, we can apply a linear transformation to the system such that the transferred system has real system matrices. The detailed description of the system models can be found in \cite{james2008h, nurdin2009coherent, zhang2010direct, zhang2012coherent}. The initial system variables $x(0) = x_0$ are Gaussian with state $\rho$, and satisfy the commutation relations:
\begin{equation}
[x_j(0), x_k(0)] = 2i \Theta_{jk}, \quad j,k=1, \dots, n,
\end{equation}
where $\Theta_{jk}$ is the element in a real antisymmetric matrix, $\Theta$. This matrix can be in one of the following forms:
\begin{itemize}
    \item Canonical, if $\Theta = {\rm diag}(J,J,...,J)$, or
    \item Degenerate canonical, if $\Theta = {\rm diag}(0_{n^{\prime} \times n^{\prime}} , J, . . . , J)$, where $0 < n^{\prime} \leq n$.
\end{itemize}
Here, $J$ denotes the real skew-symmetric $2 \times 2$ matrix:
\begin{equation}
J = \begin{bmatrix} 0 & 1 \\ -1 & 0 \end{bmatrix}.
\end{equation}
A canonical  $ \Theta $  indicates a fully quantum system, such as a set of quantum oscillators. A degenerate canonical $ \Theta $  could indicate a hybrid system where some subsystems are treated classically.

The output  $ y(t) $  corresponds to a selection of the system's output field channels, and  $ w(t) $  is a vector of input signals, decomposed as  $ dw(t) = \beta_w(t)dt + d\Tilde{w}(t) $.  $ \beta_w(t) $  is the self-adjoint adapted process, and  $ \Tilde{w}(t) $  is the noise component, a vector of self-adjoint quantum noises with the Ito table:
\begin{equation}\label{eq:noiseFmat}
\begin{aligned}
    d\Tilde{w}(t)d\Tilde{w}^{T}(t)=F_{\Tilde{w}}dt.
\end{aligned}
\end{equation}
Here,  $ F_{\Tilde{w}} $  is a nonnegative Hermitian matrix, leading to the commutation relations for the noise components:
\begin{equation}
    \begin{aligned}
    \relax [d\Tilde{w}(t),d\Tilde{w}^{T}(t)] 
    &= d\Tilde{w}(t)d\Tilde{w}^{T}(t) - (d\Tilde{w}(t)d\Tilde{w}^{T}(t))^T \\
    &= 2 T_{\tilde{w}} \, dt,
    \end{aligned}
\end{equation}
where  $ T_{\Tilde{w}} = \frac{1}{2} (F_{\Tilde{w}} - F_{\Tilde{w}}^T) $ and $S_{\Tilde{w}} = \frac{1}{2} (F_{\Tilde{w}} + F_{\Tilde{w}}^T) $, such that $ F_{\Tilde{w}} = S_{\Tilde{w}} + T_{\Tilde{w}} $ .
The form of  $ F_{\Tilde{w}} $  reflects the properties of the noise component  $ \Tilde{w} $ . For instance,  $ F_{\Tilde{w}} = \text{diag}(1, \mathbb{I} + iJ) $  signifies a noise vector with one classical component and a pair of conjugate quantum noises.

For the system described by \eqref{eq:generalSS} to be quantum, the matrices must meet the conditions for physical realizability \cite{nurdin2009coherent,james2008h,zhang2010direct,zhang2012coherent}:
\begin{equation}\label{eq:PRconditions}
\begin{aligned}
iA\Theta+i\Theta A^T+B T_{\Tilde{w}}B^T=0,\\    
B\begin{bmatrix}
    \mathbb{I}_{n_y\times n_y}\\
    0_{(n_w-n_y)\times n_y}
\end{bmatrix}=\Theta C^T{\rm diag}_{n_y/2}(J),\\
D = \begin{bmatrix}
    \mathbb{I}_{n_y\times n_y} \quad 0_{n_y\times (n_w-n_y)}
\end{bmatrix}.
\end{aligned}    
\end{equation}

\subsection{Closed-loop system}
To control the plant, a control input field is applied to the original system equation \eqref{eq:generalSS}, yielding:
\begin{equation}\label{eq:systemwithu}
\begin{aligned}
    dx(t) &= Ax(t)\,dt + B\,du(t) + B_w\,dw(t), \quad x(0) = x, \\
    dy(t) &= Cx(t)\,dt + D_w\,dw(t), \\
    z(t) &= C_z\,x(t) + D_z\,\beta_u(t).
\end{aligned}
\end{equation}
Here, $y(t)$ is the output field that will be injected to the controller to provide information on the plant. $z(t)$ is an output designed to quantify system performance metrics (e.g., error quantity, control cost). We still assume real system matrices, $A\in \mathbb{R}^{n\times n}, B\in\mathbb{R}^{n\times n_k}, B_w\in \mathbb{R}^{n\times n_w},C\in\mathbb{R}^{n_y\times n}, D_w\in\mathbb{R}^{n_y\times n_w}, C_z\in\mathbb{R}^{n_z\times n}$, $D_z\in\mathbb{R}^{n_z\times n_k}$.

The control input $u(t)$ is given by $du(t) = \beta_u(t)\,dt + d \Tilde{u}(t)$, where $\beta_u(t)$ is the signal component, and $\Tilde{u}(t)$ corresponds to the noise component. Meanwhile, $\beta_u(t)$ is an adapted, self-adjoint process that commutes with $x(t)$, satisfying the commutation relation $ \beta_u(t)x(t)^T - (x(t)\beta_u(t)^T)^T = 0 $. The vectors $w(t)$ and $\Tilde{u}(t)$ are independent quantum noise processes.

The controller is modeled as another linear quantum system, whose dynamics satisfy:
\begin{equation}
    \begin{aligned}
    d\xi(t) &= A_K \xi(t) dt + B_{K_1} dw_{K_1}(t)+ B_{K_2} dw_{K_2}(t) \\ 
    & \quad + B_{K_y} dy(t), 
    \\ du(t) &= C_K \xi(t) dt + dw_{K_1}(t),
    \end{aligned}
\label{eq:QCcontroller}
\end{equation}
where $ \xi(t) = [\xi_1(t) \cdots \xi_{n_k}(t)]^T $ is a vector of self-adjoint controller variables with the same dimension as $ x(t) $ (indicating that the controller is of the same order as the plant). The matrices $ A_K\in \mathbb{R}^{n\times n}, B_{K_1}\in\mathbb{R}^{n\times n_1}, B_{K_2}\in\mathbb{R}^{n\times n_2}, B_{K_y}\in\mathbb{R}^{n\times n_y}, C_K\in\mathbb{R}^{n_k\times n}$ are to be designed. The two additional input fields in the controller, $w_{K_1}(t), w_{K_2}(t)$, are included to ensure the physical realization condition for the quantum controller \cite{nurdin2009coherent}. In this case, after the controller matrices, $A_K, B_{K_2}, B_{K_y}, C_K$, are designed from classical control methods such as $H^\infty$ and LQG control, the matrix $B_{K_1}$ can always be constructed to ensure the physical realization conditions, which we discuss later in this section. The initial condition for the controller satisfies $ \xi(0)\xi(0)^T - (\xi(0)\xi(0)^T)^T = 2i\Theta_K$, where $ \Theta_K $ is the skew-symmetric commutation matrix for the controller variables, potentially in a canonical (quantum controller) or degenerate canonical (hybrid controller) form. We assume that $x(0)\xi(0)^T - (\xi(0)x(0)^T)^T = 0$, ensuring that the plant and controller variables commute.

The closed-loop Plant-Controller system is obtained by identifying $\beta_u(t) \equiv C_K \xi(t)$ and $\Tilde{u}(t) \equiv w_{K_1}(t)$, and interconnecting equations \eqref{eq:systemwithu} and \eqref{eq:QCcontroller}:
\begin{equation}\label{eq:closeloop}
    \begin{aligned}
    d\eta(t) &= \mathcal{A}\eta(t)dt + \mathcal{B}dw_{cl}(t), \\
    z(t) &= \mathcal{C}\eta(t),
    \end{aligned}
\end{equation}
where $ \eta(t) = [x(t)^T  \quad \xi(t)^T]^T $, and the matrices are defined as:
\begin{equation*}
w_{cl}(t) = \begin{bmatrix} w(t) \\ w_{K_1}(t) \\ w_{K_2}(t) \end{bmatrix}, \quad
\mathcal{A} = \begin{bmatrix} A & BC_K \\ B_{K_y}C & A_K \end{bmatrix},
\end{equation*}
\begin{equation*}
\mathcal{B} = \begin{bmatrix} B_w & B & 0_{2\times2} \\ B_{K_y}D_w & B_{K_1} & B_{K_2} \end{bmatrix}, \quad
\mathcal{C} = \begin{bmatrix} C_z & D_zC_K \end{bmatrix}.
\end{equation*}
Note that the matrix $\mathcal{A}$ must be Hurwitz to ensure stability.

\subsection{Control problem formulation}
\noindent The objective of control for the closed-loop system described by \eqref{eq:closeloop} is to minimize the quadratic performance index:
\begin{equation}
J(t_f) = \int_0^{t_f} \langle z^T(t)z(t)\rangle dt.
\label{qpi}
\end{equation}
Here $\langle \cdot \rangle$ is the standard quantum expectation. The quadratic form of $\langle z^T(t)z(t)\rangle$ can potentially represent the energy or magnitude of the output $z(t)$, which is a common performance index to be minimized in control theory. 
With the closed-loop system, we can rewrite the performance index to be \cite{nurdin2009coherent}:
\begin{alignat}{1}
J(t_f) &= \int_0^{t_f} \langle z^T(t)z(t)\rangle dt=\int_0^{t_f} \langle \eta^T(t) \mathcal{C}^T\mathcal{C}\eta(t)\rangle dt,\nonumber\\
&=\int_0^{t_f}{\rm Tr}(\eta^T(t)\mathcal{C}^T\mathcal{C}\eta(t))dt,\label{eq:rwp}\\
&=\int_0^{t_f}\frac{1}{2}\langle {\rm Tr}(\mathcal{C}^T\mathcal{C}[\eta(t)\eta^T(t)+(\eta(t)\eta^T(t))^T]) \rangle dt.\nonumber
\end{alignat}
We define $P(t)=\frac{1}{2}\langle \eta(t)\eta^T(t)+(\eta(t)\eta^T(t))^T \rangle$, and can derive its differential equation from the closed loop system \eqref{eq:closeloop}:
\begin{equation}
        dP(t)=\left( \mathcal{A}P(t)+P(t)\mathcal{A}^T+\mathcal{B}\mathcal{B}^T \right)dt.
\end{equation}
With this $P(t)$, we consider the infinite horizon quadratic performance index, which can be calculated as 
\begin{equation}
J_\infty =\lim_{t_f\rightarrow\infty}\sup \frac{1}{t_f}\int_0^{t_f} \langle z^T(t)z(t)\rangle dt={\rm Tr}(\mathcal{C}P\mathcal{C}^T).
\label{eq:Jinfty}
\end{equation}
Here $P=\lim_{t\rightarrow\infty} P(t)$ is the symmetric positive solution of the following Lyapunov equation:
\begin{equation}
\mathcal{A}P+P\mathcal{A}^T+\mathcal{B}\mathcal{B}^T=0.
\label{eq:pLyapunov}
\end{equation}

Therefore, we summarise our control problem as follows:\\
\textbf{Problem}: Given the plant \eqref{eq:systemwithu}, find a fully quantum controller \eqref{eq:QCcontroller} with $A_K, B_{K_1}, B_{K_2}, B_{K_y}, C_K$ to minimize $J_\infty$ \eqref{eq:Jinfty} such that:
\begin{enumerate}
    \item There exists a symmetric solution $P>0$ to the Lyapunov equation \eqref{eq:pLyapunov}.
    \item The physical constraints,    
    \begin{alignat}{1}
    &A_K\Theta_K+\Theta_K A_K^T+ B_{K_1}{\rm diag}_{n_1/2}(J)B_{K_1}^T \label{eq:PRcontroller1} \\
    &+B_{K_2}{\rm diag}_{n_2/2}(J)B_{K_2}^T+B_{K_y}{\rm diag}_{n_y/2}(J)B_{K_y}^T=0 \nonumber
    \end{alignat}
    \begin{equation}\label{eq:PRcontroller2}
        \begin{aligned} 
                B_{K_1}=\Theta_K C_K^T{\rm diag}_{n_u/2}(J)
        \end{aligned} 
    \end{equation}
    are satisfied. Here $n_1, n_2, n_y$ are the dimensions of $w_{K_1}(t), w_{K_2}(t), y(t)$, respectively.
\end{enumerate}

This quantum coherent feedback control problem has been considered in \cite{nurdin2009coherent,zhang2012coherent,zhang2010direct}. In this paper, we develop an algorithm based on DE to solve the controller design problem by transforming it into a constrained optimization problem (COP), aiming to find the optimal parameters that minimize the objective function while satisfying the constraints. Therefore, our algorithm is designed to construct the required control parameters $A_K, B_{K_1}, B_{K_2}, B_{K_y}$, and $C_K$ such that they satisfy the physical constraints \eqref{eq:PRcontroller1} and \eqref{eq:PRcontroller2} and also ensure there is a positive definite solution, $P$, to the Lyapunov function \eqref{eq:pLyapunov}. 

To reduce the algorithm computational complexity, we first consider to reduce the number of constraints. The first constraint requires that the solution $P$ to the Lyapunov equation \eqref{eq:pLyapunov} be symmetric and positive definite. This condition can be ensured by enforcing that the smallest eigenvalue of $P$, denoted as $\lambda_{\text{min}}(P)$, satisfies:
\begin{equation} \label{eq:ineq}
\lambda_{\text{min}}(P) > 0.
\end{equation}

If we define the left side of \eqref{eq:PRcontroller1} as $M$, this second constraint means that all the elements in $M$ are zeros. This makes the number of constraints to increase with the dimensions of $M$. However, we can reduce the number of constraints by only considering the largest absolute value of the matrix elements in $M$ during the algorithm searching process. Because for a feasible solution set, the largest absolute value in $M$, $\|M\|_{\mathrm{max}}=\max_{j,k}|M_{j,k}|$ satisfies:
\begin{equation} \label{eq:eq}
\|M\|_{\mathrm{max}} = 0,
\end{equation}
we can guarantee $M=0$ under this solution set.

The third constraint in \eqref{eq:PRcontroller2} can be satisfied by directly solving $B_{K_1}$ from $C_K$. By reducing the number of constraints and focusing on the key variables, the optimization process becomes more efficient while still ensuring compliance with the performance and physical realizability conditions.

We define:
\begin{equation}
    h(\vec{u}) = \lambda_{\text{min}}(P),
\end{equation}
\begin{equation} \label{eq:kx}
    k(\vec{u}) = \|M\|_{\mathrm{max}},
\end{equation} where 
\begin{equation} \label{eq:vecu}
    \vec{u} = [u(1), u(2), \cdots, u(D)]
\end{equation}
represents the decision variables to be optimized. Here $D$ is the number of total elements to be designed for the control matrices. For instance,
\begin{equation}\label{AKBK2BK3CK} 
\begin{aligned}     
\begin{bmatrix}u(1) & u(2) \\ u(3) & u(4) \end{bmatrix} = A_K, \qquad \begin{bmatrix}u(5) & u(6) \\ u(7) & u(8) \end{bmatrix} = C_K, \\     
\begin{bmatrix}u(9) & u(10) \\ u(11) & u(12) \end{bmatrix} = B_{K_2}, \qquad \begin{bmatrix}u(13) & u(14) \\ u(15) & u(16) \end{bmatrix} = B_{K_y},  
\end{aligned} 
\end{equation}
where $D=16$. The value of $k(\vec{u})$ is regarded as a measure of the degree to which the equality constraint \eqref{eq:eq} is satisfied. A larger  $k(\vec{u})$  indicates a greater violation for the constraint.
Thus, the overall optimization problem can be formulated as:
\begin{equation} \label{eq:formulatedProblem}
\begin{aligned}
\text{Minimize:} \quad & \mathcal{O}(\vec{u}) = J_\infty, \\
\text{Subject to:} \quad & h(\vec{u}) = \lambda_{\text{min}}(P) > 0, \\
& k(\vec{u}) = \|M\|_{\mathrm{max}} = 0.
\end{aligned} 
\end{equation}

In this formulation, the function $\mathcal{O}(\vec{u})$ represents the objective function to minimize, $h(\vec{u})$ ensures the positive definiteness of the matrix $P$ through inequality constraint \eqref{eq:ineq}, and $k(\vec{u})$ enforces the equality constraint \eqref{eq:eq}. Now we have formulated the coherent control design problem into a COP, which will be solved in the following section using the tailored differential evolution algorithm.

\section{Tailored Differential Evolution Algorithm for Quantum Coherent Feedback Control}
\label{sec:QCTDE}

\subsection{Differential Evolution}
\label{DE}
\noindent The DE algorithm \cite{storn1997differential,storn1996usage,price2006differential,wang2021multiobjective} has found widespread application in various scientific and engineering fields, dealing with the challenge of identifying the optimal solutions under specific constraints \cite{das2010differential,opara2019differential}. The typical framework for these problems is to find solutions that optimize a predetermined objective function subject to constraints. To measure how close we are to this ``optimal'' solution, a fitness function needs to be defined for the problem.

In this paper, as described in \eqref{eq:formulatedProblem}, our optimization objective is to determine the parameter vector $\vec{u}$ of the coherent LQG quantum controller that not only minimizes the system performance index $J_\infty$ defined in \eqref{eq:Jinfty}, but also complies with the unique constraints of the quantum system.

\begin{figure*}[htbp]
\centerline{\includegraphics[width=1\textwidth]{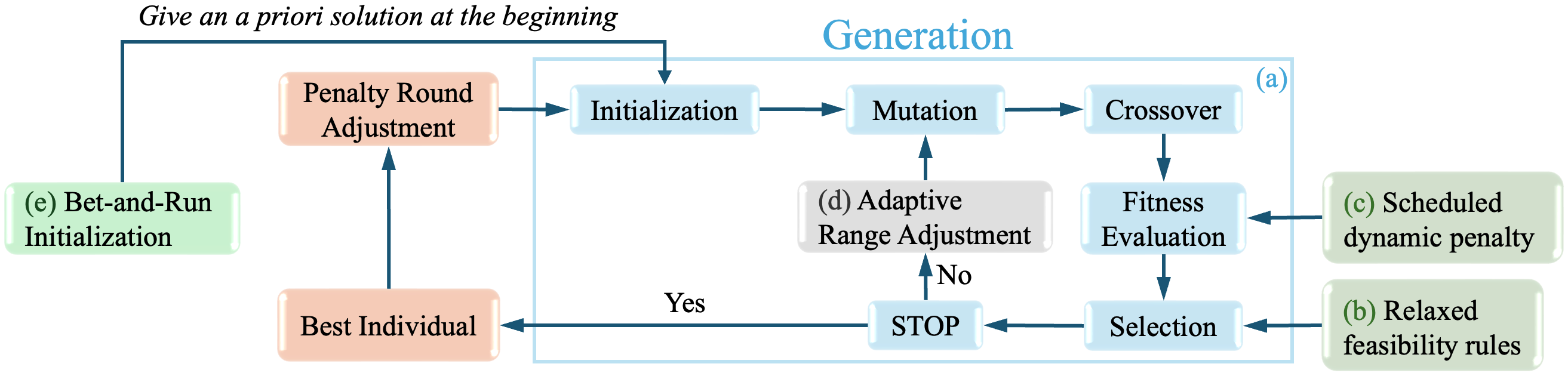}}
\caption{The workflow of the proposed algorithm. It starts with a ``bet and run'' initialization to select a promising solution for further refinement. In each $p$ round of the penalty term, the complete DE algorithm is executed, including ``initialization'', ``mutation'', ``crossover'', and ``selection''. The process combines ``adaptive range adjustment'' to efficiently explore the search space, while integrating ``relaxed feasibility rules'' and ``scheduled dynamic penalty'' to balance performance optimization and constraint satisfaction. The optimal individual in the final round is our solution.}
\label{Pic:OurDE}
\end{figure*}

DE, as an evolutionary algorithm, operates on a population 
\begin{equation} \label{eq:QCpopulation}
    \textbf{U}_g = [\vec{u}_{g,1}, \vec{u}_{g,2}, \cdots, \vec{u}_{g,Np}],
\end{equation} 
where  $g$  denotes the generation number. In each generation, the population consists of  $N_p$  individuals. Each individual  $\vec{u}_{g,i}$  is represented as an  $D$-dimensional real-valued vector, commonly referred to as a genome,
\begin{equation} \label{eq:QCgenome}
    \vec{u}_{g,i} = [u_{g,i}(1), u_{g,i}(2), \cdots, u_{g,i}(D)].
\end{equation} 
In this context, each genome $\vec{u}_{g,i}$ corresponds to a potential configuration of the quantum controller's parameters $\vec{u}$ \eqref{eq:vecu}. The basic structure of DE algorithm is shown in the blue square labeled as part (a) in Fig.~\ref{Pic:OurDE}, which is comprised of initialization, mutation, crossover, fitness evaluation, and selection. DE iteratively refines these genomes, leading to improved solutions over successive generations.
The main components of the standard DE algorithm are briefly described in the Appendix. In this paper, we propose a new algorithm based on the foundation of the standard DE framework, integrating the following improvements:
\begin{itemize}
    \item Relaxed feasibility rules: Feasibility rules are commonly used in constrained optimization problems (COPs) to limit the search space of the algorithm to solutions that satisfy the constraints \cite{mezura2011constraint}. However, a strict feasibility rule may impose too stringent constraints, making it difficult for the algorithm to find good initial solutions, thus affecting the quality of the final solution \cite{lagaros2023constraint}. We adopt a relaxed feasibility rule that allows individuals to violate the constraints to some extent, thus enhancing the diversity of the search space.
    \item Scheduled dynamic penalty function: Although relaxed feasibility rules increase the diversity of the search space, they inevitably produce solutions that do not satisfy the desired constraints. To address this concern, we introduce a scheduled dynamic penalty function. In the early stages, the function imposes a low penalty on individuals that violate the constraints, with the penalty increasing over time. This approach enables the algorithm to explore a wider range of solutions during the initial phase, gradually guiding it towards the feasible region.
    \item Adaptive range adjustment: By conducting searches within a smaller range and subsequently restoring the search values for the final application, this algorithm can more effectively handle problems with significant variations in magnitude; The adaptive search range, which is dynamically adjusted based on the current optimal solution, ensures a balance between global exploration and local exploitation. This approach avoids the challenges associated with defining a fixed search space when the solution distribution is unknown, thereby enhancing the probability and efficiency of finding the optimal solution.
    \item Bet-and-Run initialization: Prior to the formal start of the algorithm, multiple short exploratory runs are conducted to select a promising initial solution for the optimization process. This method increases the likelihood of early convergence to an optimal solution by providing a favorable starting point in the search space.
\end{itemize}
These elements are critical to achieve the controller design under the constraints. In the following we explain these elements in details.

\subsection{Relaxed Feasibility Rules}
\label{sec:RelaxedFeasibilityRules}
\noindent In the context of control systems, feasibility is typically characterized by the system's adherence to all imposed constraints. As shown in block $(b)$ of Fig.~\ref{Pic:OurDE}, the feasibility rules guide the algorithm's selection process, ensuring that decisions prioritize solutions that respect the constraints. The selection of solutions follows these principles \cite{deb2000efficient,wang2015incorporating}:
\begin{itemize}
    \item \textit{Priority of Feasible Solutions}: A feasible solution is always preferred over an infeasible one, regardless of objective function performance.
    \item \textit{Comparison Among Infeasible Solutions}: The solution with fewer or less severe constraint violations is favored.
    \item \textit{Comparison Among Feasible Solutions}: Feasible solutions are ranked by their objective function values.
\end{itemize}

For the COP addressed in this work, the degree of constraint violation for a decision vector $\vec{u}$ \eqref{eq:vecu} is computed as:
\begin{equation} \label{eq:CVdegree}
    V(\vec{u}) = \max\{0, -h(\vec{u}) + \phi\} + \max\{0, k(\vec{u}) - \delta\},
\end{equation}
which is used to measure the extent to which the solution $\vec{u}$ violates the constraints. 
Here, two tunable parameters, $\phi$ and $\delta$, are used to quantify the extent to which the two constraints in \eqref{eq:formulatedProblem} are violated. For example, $V(\vec{u}) = 0$ implies that the constraints are fully satisfied. In this case, a small positive value for $\phi$ ensures that $h(\vec{u})$ remains positive for feasible solutions from the first term in \eqref{eq:CVdegree}, while a small positive tolerance, $\delta$, ensures that $k(\vec{u}) \leq \delta$ from the second term in \eqref{eq:CVdegree}.
$k(\vec{u})$ represents the maximum absolute value of a matrix, and we have $0\leq k(\vec{u})\leq\delta$. By choosing a small $\delta$ we can let the constraint $k(\vec{u})\rightarrow 0$. Otherwise, $V(\vec{u}) \neq 0$ defines an infeasible solution $\vec{u}$. With the value of $V(\vec{u})$, the decision space of the COP is divided into a feasible region and an infeasible region.

With the value in \eqref{eq:CVdegree}, each candidate solution is evaluated based on both its objective function value and the degree of its constraint violation. Feasible solutions are given priority, and in cases where no feasible solution exists, the solution with the smallest degree of constraint violation is chosen. This approach drives the population toward the feasible region while simultaneously optimizing the objective function.

In principle, the parameter $\delta$ should be set as small as possible, ideally approaching $0$, to ensure that the equality constraint \eqref{eq:eq} be strictly satisfied. However, enforcing such strict constraints often leads to local optimization, and in severe cases, it can prevent the algorithm from finding feasible solutions, especially in complex systems where exploration of the solution space is limited \cite{wang2015incorporating,price2006differential}. In this study, we adopt a relaxed feasibility rule by setting $\delta$ to a relatively large value, such as $0.1$ or $0.01$. This adjustment broadens the search space and allows the initial optimization to focus on the objective function while still satisfying the inequality constraints \eqref{eq:ineq}. However, this relaxation in feasibility rule may result in the final value of \eqref{eq:kx} oscillating around $\delta$, which conflicts with the stringent physical constraints in \eqref{eq:eq}. To address this issue, the relaxed feasibility rule is applied alongside a scheduled dynamic penalty scheme, introduced in the following section, to gradually steer the solution toward achieving full physical realizability.

\subsection{Scheduled Dynamic Penalty}
\label{sec:dynamicPenalty}
\noindent In order to solve problems related to constraint relaxation, we propose a penalization module that is integrated into the fitness function to help the selection procedure (see part (c) of Fig.~\ref{Pic:OurDE}). Infeasible solutions are penalized but not immediately discarded, as they may be close to the optimal feasible region. As the evolution proceeds, the final solution satisfies the requirements of the quantum system while optimizing performance.

The fitness function integrating the penalty function is defined as follows:
\begin{equation} \label{fitness2}
\textit{fitness} := f(\vec{u}) = J_\infty + \mathfrak{p} \times \varrho,
\end{equation}
where $J_\infty$ denotes the objective function to be minimized, as described in \eqref{eq:Jinfty}, and $\mathfrak{p}$ represents the penalty term designed to enforce constraint satisfaction. Specifically, in our paper, $\mathfrak{p}$ is formulated as 
\begin{equation}
    \mathfrak{p} = k(\vec{u}),
\end{equation}
to represent the penalty on the constraint $k(\vec{u})$. Here, we have $\mathfrak{p} = 0$ when constraint \eqref{eq:eq} is fully satisfied; conversely, the value of $k(\vec{u})$ is taken. This penalty term serves to steer the optimization process towards feasible regions of the solution space by discouraging violations of the physical realizability constraints. The parameter $\varrho$ acts as a penalty scaling factor, regulating the relative influence of the penalty term $\mathfrak{p}$ on the overall fitness value. By tuning $\varrho$, we control the trade-off between minimizing the objective function $J_\infty$ and achieving constraint satisfaction. 

The choice of $\varrho$ plays a crucial role in shaping the optimization process. A fixed penalty factor, where $\varrho$ remains constant throughout the optimization, is commonly used \cite{michalewicz2013genetic,saha2014fuzzy}. However, this approach may lack flexibility across diverse problems, leading to challenges in striking a balance between performance and constraint satisfaction. If the penalty weight is set too high, the algorithm may prematurely converge on feasible solutions, curbing exploration near the boundary of feasible regions and leading to suboptimal performance in terms of objective function. Conversely, a penalty that is too low may result in prolonged exploration within infeasible regions, where the penalty's influence becomes negligible relative to the objective function and an infeasible solution may be found \cite{riche1993optimization}.

In this paper, we propose a scheme called ``Scheduled Dynamic Penalty'', where a non-stationary $\varrho$ is used. This scheme can adaptively increase the value of $\varrho$ over generations, to encourage convergence toward feasible solutions throughout the optimization process \cite{joines1994use}.
Initially, the search space is expanded to explore promising potential solutions, while the penalty factor $\varrho$ is dynamically adjusted to increasingly penalize solutions that deviate from the constraints. This approach gradually shifts the optimization towards physically realizable results. By balancing early-stage exploration with later-stage constraint enforcement, it ensures flexibility in the initial search while steering the algorithm toward feasible solutions over time.

\begin{figure}[htbp]
\centerline{\includegraphics[width=0.48\textwidth]{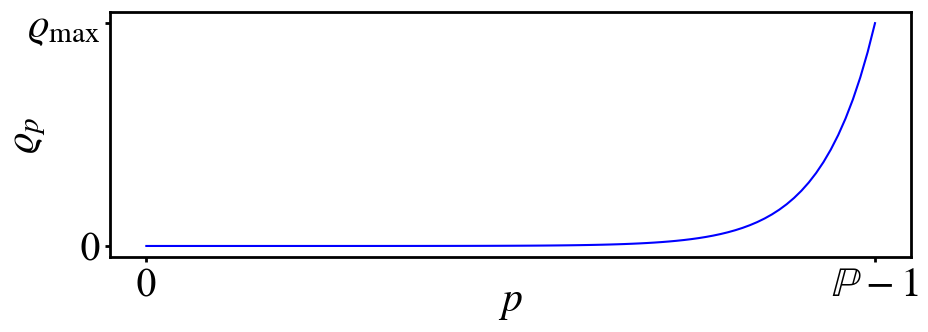}}
\caption{The penalty factor $\varrho_p$ increases with the step number $p$. In the early stages, only a small penalty is applied to infeasible solutions, with an emphasis on improving system performance. In the later stages, the penalty for violating constraints increases rapidly, aiming to find feasible solutions near the region of solutions with good performance.}
\label{pic:penaltyFactor}
\end{figure}

In our framework, the penalty factor is chosen as:
\begin{equation}
\varrho_p = (\varrho_{\text{max}})^{\frac{p}{\mathbb{P}-1}},
\end{equation}
where $p = 0, 1, 2, \dots, \mathbb{P}-1$, $\mathbb{P}$ is the total number of penalty steps (also how many different penalty factors being used), and $\varrho_{\text{max}}$ regulates the precision of constraint satisfaction. As shown in Fig.~\ref{pic:penaltyFactor}, the penalty factor $\varrho_p$ grows more rapidly as the step number $p$ increases. By adjusting $\mathbb{P}$, we can make a trade-off between computational cost and the finesse of the solution. The larger the value of $\mathbb{P}$, the finer the control of the optimization process, and the more likely it is to result in a better solution, but with an increased computational burden. Similarly, adjusting the value of $\varrho_{\text{max}}$ controls the degree to which equality constraints in the physical realizability requirements are enforced, with larger values indicating a tighter emphasis on constraint precision. 
We refer to each $p$ as a ``round''. This scheduled approach ensures that the penalty factor increases smoothly with each round, thus gradually shifting the focus of optimization from simply improving performance to balancing both performance and constraints.

For the $p$ round, the DE algorithm runs to completion, retaining the best solution from the final generation as the prior initial solution for the $p+1$ round. The updated population and genomes for penalty step $p$ are represented as:
\begin{equation} \label{penaltyPopulation}
\textbf{U}^p_g = [\vec{u}^p_{g,1}, \vec{u}^p_{g,2}, \cdots, \vec{u}^p_{g,Np}],
\end{equation}
\begin{equation} \label{penaltyGenome}
\vec{u}^p_{g,i} = [u^p_{g,i}(1), u^p_{g,i}(2), \cdots, u^p_{g,i}(D)],
\end{equation}
where genome $\vec{u}^p_{g,i}$ represents the $i\text{-th}$ candidate solution of generation $g$ in the $p\text{-th}$ round. At each $p$ round, the fitness function is defined as:
\begin{equation} \label{fitness3}
    f(\vec{u}) = J_\infty + \mathfrak{p} \times \varrho_p.
\end{equation}
Again, $J_\infty$ represents the performance objective, and $\mathfrak{p}$ quantifies the violation of constraint \eqref{eq:eq}. Initially, when $\varrho_p$ is small, the algorithm prioritizes minimizing $J_\infty$, allowing for exploration of the search space. As $\varrho_p$ increases, the penalty term becomes more significant, encouraging solutions that satisfy both performance and feasibility requirements.

This approach combines relaxed feasibility rules with a scheduled dynamic penalty method to provide a robust framework for the constrained optimization problem of quantum coherent feedback controllers, by balancing the dual goals of minimizing $J_\infty$ and satisfying physical constraints. The relaxed feasibility rules help prevent search difficulties caused by overly stringent constraints, while also reducing the search space to some extent. By iteratively adjusting the penalty factor, the method ensures that the final solution is both optimal and feasible, thus effectively addressing the complexity of constraint satisfaction in quantum coherent control problems.

\subsection{Adaptive Range Adjustment}
\label{sec:SearchRangeAdjustment}
\noindent In this section, we introduce a scheme to dynamically change the search space for the DE algorithm, based on the real-time solutions. The core idea includes two main parts, search value scaling and adaptive search range. As shown in part (d) of Fig.~\ref{Pic:OurDE}, this module is used during initialization, mutation, and crossover.

\subsubsection{Search Value Scaling}
\label{sec:SearchValueScaling}
\noindent This method transforms the search space of all variables into a smaller, unified range, for better exploitation, then scales the solutions back to the values for the actual application. Therefore, this method is effective for problems involving variables with large magnitude disparities, as direct searches in very large or very small ranges may lead to inefficiency. This transformation is governed by the equation:
\begin{equation}
\vec{u}^p_{g,i} = \alpha \cdot \vec{u}^{\prime p}_{g,i}. 
\end{equation}
Here, $\vec{u}^{\prime p}_{g,i}$ represents the values during the search process, $\vec{u}^{p}_{g,i}$ denotes the actual values applied to the system design (also used to calculate the fitness function). $\alpha$ is the scaling factor, which is greater than or equal to $1$. When $\alpha = 1$, the values obtained from the search are directly used as system parameters; otherwise, the values are scaled by a factor of $\alpha$ before being applied to the system.

The scaled search space method possesses several notable advantages, particularly in the absence of prior knowledge and when dealing with optimization problems where variables exhibit large differences in magnitude. Conducting searches on a unified scale enhances the precision and numerical stability of the optimization process, as it better balances and fine-tunes variables across different orders of magnitude, thereby reducing the risk of numerical instability that may arise from optimizing extremely large and small values without appropriate scaling. Additionally, this approach increases the likelihood of finding a good initial solution, as confining the search to a smaller range significantly improves the chances of converging to a feasible starting point, which can then be scaled to the required magnitude.

\subsubsection{Adaptive Search Range}
\label{sec:AdaptiveSearchRange}
\begin{figure}[htbp]
\centerline{\includegraphics[width=0.48\textwidth]{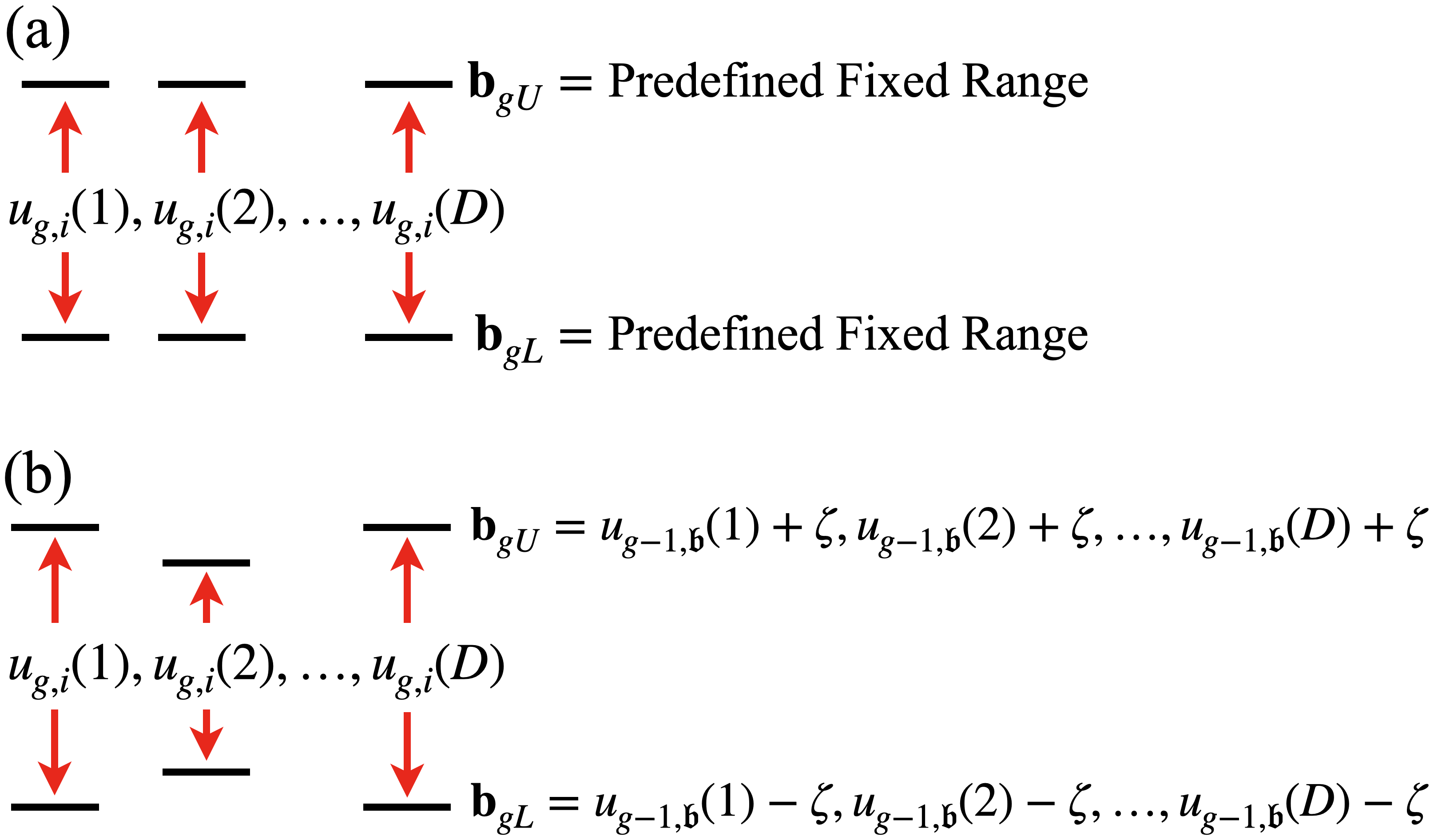}}
\caption{Search Range. (a) Fixed search range; (b) Adaptive search range.}
\label{searchRange}
\end{figure}

\noindent The second idea in this adaptive range adjustment is called adaptive search range. Traditionally, the search range for each variable is fixed, and these boundary values are represented as two D-dimensional vectors:
\begin{equation} \label{eq:QCfixedBounds}
\begin{aligned}
    \mathbf{b}_L = [b_{L1}, b_{L2}, \ldots, b_{LD}], \\ 
    \mathbf{b}_U = [b_{U1}, b_{U2}, \ldots, b_{UD}].  
\end{aligned}
\end{equation}
Here, the subscripts $L$ and $U$ refer to the lower and upper bounds, respectively. Fig.~\ref{searchRange} (a) illustrates this fixed-boundary approach, where the upper and lower bounds are often set symmetrically around zero and uniformly across variables, in the absence of specific requirements.

Although this fixed range ensures consistency, it can lead to inefficiency, especially when the optimal values for different variables are located in vastly different regions.
To address this limitation, we propose an adaptive search range strategy, as shown in Fig.~\ref{searchRange} (b), where the search boundaries for each variable are dynamically adjusted in every generation, allowing the algorithm to focus its exploration on promising regions determined by the current best solution. Specifically, as illustrated in part (d) of Fig.~\ref{Pic:OurDE}, at the end of generation $(g-1)$, we use the current best solution as a reference to adjust the search boundaries for the next generation $(g)$:
\begin{equation}
\mathbf{b}_{gL} = [u_{g-1,\mathfrak{b}}(1) - \zeta, u_{g-1,\mathfrak{b}}(2) - \zeta, \dots, u_{g-1,\mathfrak{b}}(D) - \zeta]
\end{equation}
\begin{equation}
\mathbf{b}_{gU} = [u_{g-1,\mathfrak{b}}(1) + \zeta, u_{g-1,\mathfrak{b}}(2) + \zeta, \dots, u_{g-1,\mathfrak{b}}(D) + \zeta]
\end{equation}
where $u_{g-1,\mathfrak{b}}(j)$ represents the $j$-th component of the best solution from the previous generation, and $\zeta$ controls the magnitude of the adjustment. Larger values of $\zeta$ promote broader exploration but may slow down convergence, while smaller values enhance exploitation by narrowing the search to finer regions.

The scaling method ensures numerical stability and efficient exploration by transforming variables into manageable ranges, while the adaptive range method dynamically adjusts search boundaries to focus computational effort on promising regions. These strategies combined together to significantly improve the robustness, precision, and efficiency of the optimization process, particularly for problems with complex search spaces.

\subsection{Bet-and-Run Initialization}
\label{sec:BRI}
\noindent DE is generally regarded as insensitive to the choice of initial values, with algorithms typically beginning from random solutions within the defined search space \cite{kenneth2006evolutionary}. However, selecting a strategically advantageous starting point can substantially accelerate convergence and improve the algorithm's overall performance in locating the global optimum \cite{das2010differential,ali2013unconventional}. In this work, we incorporate a variation of the ``bet-and-run'' strategy \cite{fischetti2014exploiting}, a method originally developed for solving combinatorial optimization problems.

The ``bet-and-run'' approach consists of two phases. Initially, several short preliminary runs are performed from different random initial conditions, each for a limited number of iterations or evaluations. Based on a predefined selection criterion, such as the quality of the solution obtained within these short runs, the most promising candidate is selected. It serves as the very first step in the formal running of our algorithm, as shown in part (e) of Fig.~\ref{Pic:OurDE}.  This selected candidate is then subjected to a full optimization process, benefiting from the prior exploratory phase.

In our approach, we apply relaxed feasibility rules (see Section~\ref{sec:RelaxedFeasibilityRules}) to relax the constraints and thus explore a wider solution space. We start $I$ independent DE runs with random starting points, and the initial search range for all parameters is set to $[-b, b]$. Using the search value scaling and adaptive range adjustment discussed in the previous section, after $N_B$ iterations, the solution with the smallest objective function value is selected as the starting point for the subsequent optimization process. While this initial solution may not fully satisfy the constraints, it provides a promising starting position, increasing the probability of converging to an optimal and physically realizable controller.

\subsection{Early Termination}
\noindent For each $p$ round, the algorithm is configured to evolve for up to $300,000$ generations. To monitor stagnation in the optimization process, we define a convergence threshold of $10^{-10}$. This threshold represents the minimum change in fitness required for progress. If the improvement in fitness between consecutive generations is less than this threshold, the algorithm is considered to have reached stagnation. In addition, a stagnation counter is used, which is capped at $60,000$ generations. This means that for a given penalty factor $\varrho_p$, the evolutionary process will either continue for $300,000$ generations or terminate prematurely when the stagnation condition is met. In this case, the algorithm moves to the next round $p+1$. This early termination mechanism helps prevent unnecessary computations and speeds up the transition to the next optimization stage.

The key flow of the DE algorithm incorporating multiple optimization modules is shown in Fig.~\ref{Pic:OurDE}.

\section{Applications to quantum systems}\label{sec:Apps}
\noindent In this section, we apply the designed algorithm to a linear quantum optical system for LQG quantum controller design, where three different controllers are designed representing three closed-loop system configurations.

\begin{figure*}[htbp]
\centerline{\includegraphics[width=1\textwidth]{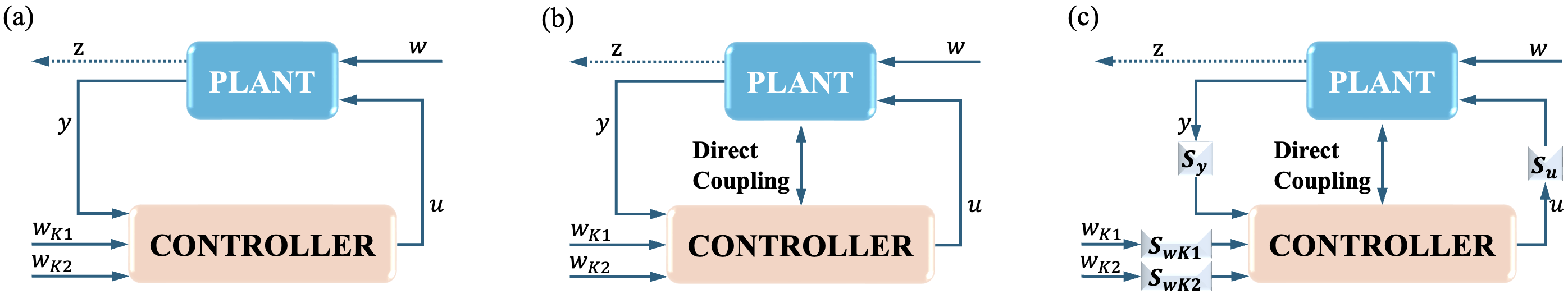}}
\caption{Applications (a) Indirect Coupling: The plant and controller interact indirectly through a shared external medium, such as a quantum field, without direct energy transfer between the systems;
(b) Combined Indirect and Direct Coupling: The system includes both indirect interactions via an external medium and direct energy exchanges, enabling simultaneous coupling methods between the plant and controller;
(c) Indirect and Direct Coupling with Ideal Squeezers: The system integrates both indirect and direct coupling mechanisms, with ideal squeezers compressing input signals before they are introduced into the systems, optimizing signal processing for enhanced control.}
\label{ModelAll}
\end{figure*}

The three different configurations include:
\begin{enumerate}
\item A controller with direct coupling to the plant, as shown in Fig.~\ref{ModelAll}~(a);
\item A controller with both indirect and direct couplings to plant, as shown in Fig.~\ref{ModelAll}~(b);
\item A controller with both indirect and direct couplings to the plant, and also including squeezed fields at the input to both the plant and the controller, as shown in Fig.~\ref{ModelAll}~(c).
\end{enumerate}
In all cases, the controller is fully quantum, allowing quantum information to be processed directly within the feedback loop without the need for classical measurement. This approach avoids the disturbances or information loss typically associated with classical measurement processes. Our objective is to design a fully quantum controller that minimizes the performance index $J_\infty$ \eqref{eq:Jinfty}.

The quantum plant we consider is an atom being trapped in a three-mirror cavity. This experimental setup is commonly used to continuously measure the position of the atom. The experimental protocol leverages the off-resonant interaction between the trapped atom and the cavity field to continuously monitor the atomic position via homodyne detection of the output light's phase quadrature \cite{doherty1999feedback,gough2007singular}. Under the assumption of strong coupling limit the cavity dynamics are adiabatically eliminated, thereby only leaving the dynamics of the atom to be described by \cite{nurdin2009coherent,zhang2010direct,zhang2012coherent}: 
\begin{equation}
    \begin{aligned}
    dx(t) &= \begin{bmatrix}0 & \triangle \\ -\triangle & 0 \end{bmatrix}xdt + \begin{bmatrix}0 & 0 \\ 0 & -2\sqrt{k_1} \end{bmatrix}du 
    \\ & \quad + \begin{bmatrix}0 & 0 & 0 & 0 \\ 0 & -2\sqrt{k_2} & 0 & -2\sqrt{k_3} \end{bmatrix} \begin{bmatrix}dw_1 \\ dw_2 \end{bmatrix};  \\
    dy(t) &= \begin{bmatrix}2\sqrt{k_2} & 0 \\ 0 & 0 \end{bmatrix}xdt + dw_1,
    \end{aligned}
\label{eq:plant_e1}
\end{equation}
where $\triangle = 0.1$ and $k_1 = k_2 = k_3 = 10^{-2}$. 
To asymptotically stabilize the system \eqref{eq:plant_e1} using another quantum system as the LQG controller, we set $z = x + \beta_u$; that is, $C_z = \mathbb{I}_{2\times 2} = D_z$.

\subsection{Indirect Coupling}
\label{sec:subModelA}
\noindent The first controller configuration we consider is shown in Fig.~\ref{ModelAll}~(a), where the designed controller couples with the plant via the only shared fields $y$ and $u$. This setup is referred to as indirect coupling \cite{nurdin2009coherent}. In this configuration, the plant is represented by \eqref{eq:systemwithu}, and the controller is represented by \eqref{eq:QCcontroller}. Using the method proposed in Section~\ref{sec:QCTDE}, we determine the vector $\vec{u}$ in \eqref{eq:formulatedProblem}, which here corresponds to the controller parameters $A_K$, $B_{K_2}$, $B_{K_y}$, and $C_K$, (the parameter $B_{K_1}$ is obtained by substituting $C_K$ into \eqref{eq:PRcontroller2}) to satisfy the constraints \eqref{eq:ineq} and \eqref{eq:eq}, while minimizing the performance index $J_\infty$ as defined in \eqref{eq:Jinfty}.

Nurdin et al. \cite{nurdin2009coherent} formulated the quantum LQG problem as a rank-constrained LMI problem and proposed a numerical procedure based on an alternating projection algorithm to obtain a fully quantum LQG controller. The resulting controller achieved a performance index of $J_\infty = 4.1793$ and a constraint satisfaction metric of $k(\vec{u}) = 6.098 \times 10^{-14}$.

\begin{figure}[htbp]
\centerline{\includegraphics[width=0.49\textwidth]{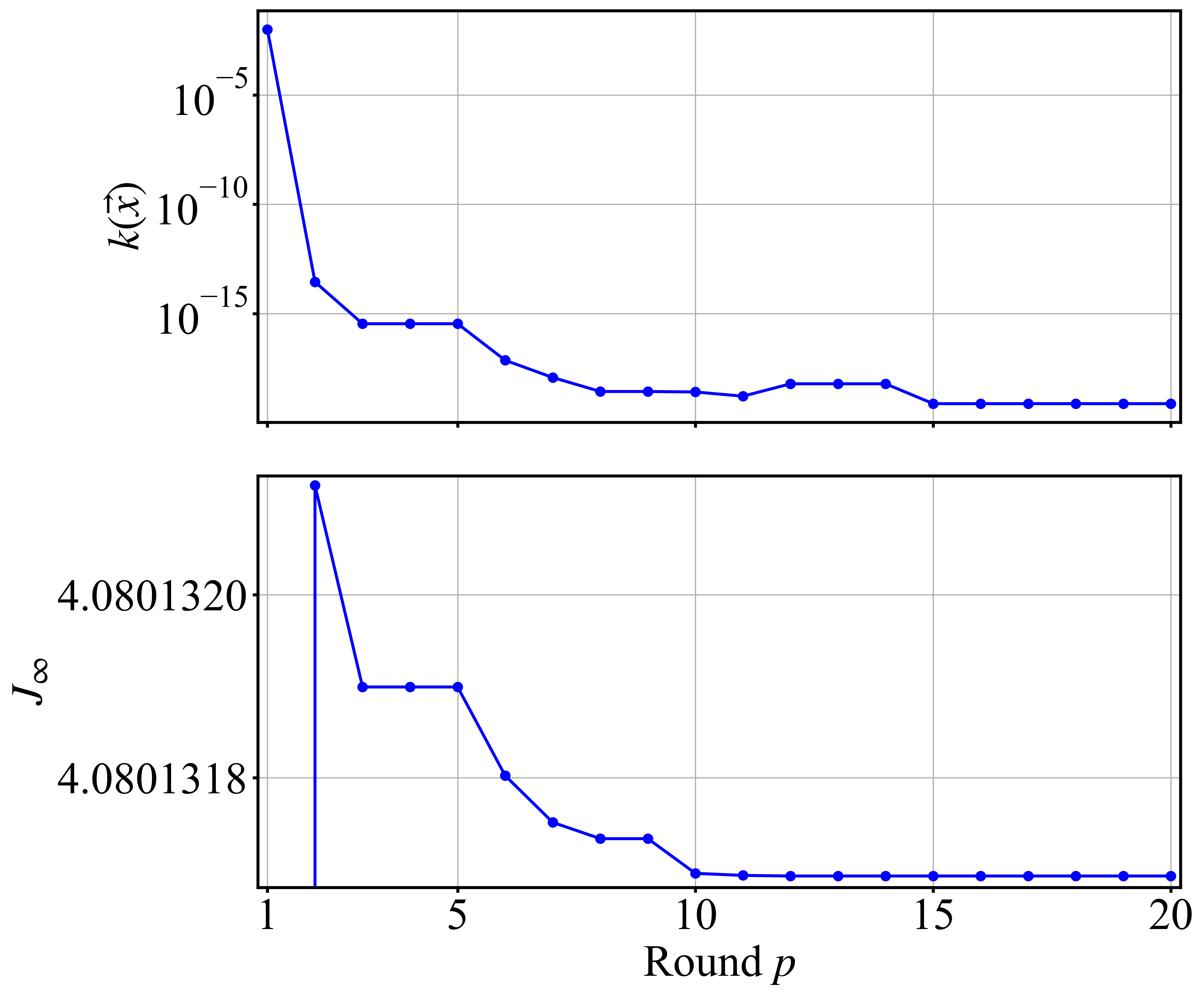}}
\caption{Enhanced Differential Evolution for Indirect Coupling Model in Section~\ref{sec:subModelA}}
\label{CombinedPic_Hendra}
\end{figure}

In the implementation of our proposed algorithm, the following parameters are configured: the parameter $\delta$ in the relaxed feasibility rules in Section~\ref{sec:RelaxedFeasibilityRules} is set to $0.01$. The number of penalty steps in the scheduled dynamic penalty module in Section~\ref{sec:dynamicPenalty} is set to $P=20$, with an accuracy factor of $\varrho_{\text{max}}=10^{10}$. In the search value scaling in Section~\ref{sec:SearchValueScaling}, the scaling factor is set to $\alpha=10$. For the adaptive search range module in Section~\ref{sec:AdaptiveSearchRange}, the adjustment magnitude for the dynamic search boundaries is set to $\zeta=1$. Additionally, within the Bet-and-Run Initialization framework in Section~\ref{sec:BRI}, the number of independent runs is configured as $I=10$, the number of iterations is $N_B=10$, with the initial search range defined as $[-b, b] = [-1, 1]$. As a result, this scheme generated the following optimized controller:
\begin{alignat}{1}
 d\xi & = \begin{bmatrix}-0.16276908 & -0.89563025 \\
 0.01098282 & -0.05451424 \end{bmatrix} \xi dt \nonumber 
 \\ & \quad + \begin{bmatrix}0.06534092 & 1.09216533 \\
 -0.09948915 & -0.01421346 \end{bmatrix} dw_{K_1} \nonumber 
 \\ & \quad + 10^{-5} \begin{bmatrix}2.17696660 & -6.37881646 \\
 -0.46006743 & -0.27001274 \end{bmatrix} dw_{K_2} \nonumber 
 \\ & \quad + \begin{bmatrix}-0.16454671 & 0.39440638 \\
 -0.27727833 & -0.00117352\end{bmatrix} dy;
 \\ du(t) &= \begin{bmatrix}0.01421346 & 1.09216533 \\
 -0.09948915 & -0.06534092 \end{bmatrix} \xi dt + dw_{K_1}.\nonumber 
\end{alignat}
As shown in Fig.~\ref{CombinedPic_Hendra}, as the number of penalty rounds $p$ increases, the controller gradually goes from focusing only on the performance index $J_\infty$ at the beginning, to balancing constraints and performance, and finally achieves a performance index of $J_\infty = 4.08013169$, and a constraint satisfaction index of $k(\vec{u}) = 7.66610068 \times 10^{-20}$. This represents a significant improvement over the existing LQG control scheme \cite{nurdin2009coherent}, both in terms of physical realizability ($7.66610068 \times 10^{-20}  < 6.098 \times 10^{-14}$) and performance index ($4.08013169 < 4.1793$). The remarkably small value of $k(\vec{u})$ indicates that our controller adheres to the physical realizability constraints with high precision while also achieving superior performance.

\subsection{Direct and Indirect Couplings}
\label{sec:subModelB}
\noindent In the previously considered application in Section~\ref{sec:subModelA}, the plant \eqref{eq:systemwithu} and the controller \eqref{eq:QCcontroller} are coupled only indirectly. We now explore a scenario where both indirect and direct couplings are present between the plant and the controller to further enhance performance. As shown in Fig.~\ref{ModelAll} (b), direct coupling is a physical mechanism in which quantum devices and their quantum controllers can exchange energy directly, without requiring a Boson field connection\cite{zhang2010direct}. The dynamics of the plant is described by:
\begin{equation}
    \begin{aligned}
    dx(t) &= Ax(t)dt + B_{12}d\xi(t) + Bdu(t) + B_w dw(t), \\ 
    x(0) &= x, \\
    dy(t) &= Cx(t)dt + D_w dw(t), \\
    z(t) &= C_z x(t) + D_z \beta_u(t),
    \end{aligned}
\label{eq:QCplantB}
\end{equation}
where the non-zero matrix, $B_{12}$, directly couples the controller to the plant. The plant in \eqref{eq:plant_e1} then becomes:
\begin{alignat}{1}
    dx(t) &= \begin{bmatrix}0 & \triangle \\ -\triangle & 0 \end{bmatrix}xdt + B_{12}d\xi(t) + \begin{bmatrix}0 & 0 \\ 0 & -2\sqrt{k_1} \end{bmatrix}du \nonumber\\ 
    & \quad + \begin{bmatrix}0 & 0 & 0 & 0 \\ 0 & -2\sqrt{k_2} & 0 & -2\sqrt{k_3} \end{bmatrix} \begin{bmatrix}dw_1 \\ dw_2 \end{bmatrix};  \label{plant_e2} \\
    dy(t) &= \begin{bmatrix}2\sqrt{k_2} & 0 \\ 0 & 0 \end{bmatrix}xdt + dw_1.  \nonumber
\end{alignat}
This direct coupling will also modify the controller's model to
\begin{equation}
    \begin{aligned}
    d\xi(t) &= A_K \xi(t) dt + B_{21}dx(t) + B_{K_1} dw_{K_1}(t) \\ 
    & \quad + B_{K_2} dw_{K_2}(t) + B_{K_y} dy(t), \\
    du(t) &= C_K \xi(t) dt + dw_{K_1}(t).
    \end{aligned}
\label{eq:QCcontrollerB}
\end{equation}
Consequently, the matrices $\mathcal{A}$, $\mathcal{B}$, and $\mathcal{C}$ in the closed-loop system \eqref{eq:closeloop} are modified as follows:
\begin{equation*}
\mathcal{A} = \begin{bmatrix} A & BC_K + B_{12} \\ B_{K_y}C + B_{21} & A_K \end{bmatrix};
\end{equation*}
\begin{equation*}
\mathcal{B} = \begin{bmatrix} B_w & B & 0_{2\times2} \\ B_{K_y}D_w & B_{K_1} & B_{K_2} \end{bmatrix}; \quad
\mathcal{C} = \begin{bmatrix} C_z & D_zC_K \end{bmatrix}.
\end{equation*}

The controller \eqref{eq:QCcontrollerB} is considered physically realizable if, in addition to satisfying the constraints \eqref{eq:ineq} and \eqref{eq:eq}, it also meets the additional requirement \cite{zhang2010direct}:
\begin{equation}
    B_{21} = \Theta_K B_{12}^T {\rm diag}_{n_u/2}(J).
    \label{eq:QCconstraint3}
\end{equation}

Zhang et al. \cite{zhang2010direct} proposed a multi-step optimization algorithm, which, based on the controller in \cite{nurdin2009coherent}, incorporates direct couplings $B_{12}$ and $B_{21}$. This approach improves the performance index while maintaining the same level of equality constraints ($k(\vec{u}) = 6.098 \times 10^{-14}$), achieving $J_\infty = 4.000049633093338$. This performance outperforms the LQG controller proposed in \cite{nurdin2009coherent} ($J_\infty = 4.1793$).

Our designed algorithm can also be applied to the controller design when considering this direct coupling configuration. We apply the approach presented in Section~\ref{sec:QCTDE} to this task and set the same algorithmic parameters as in the previous Section~\ref{sec:subModelA}. In addition to designing the controller parameters, it involves determining the direct coupling parameters  $B_{12}$  and $B_{21}$, which introduces an additional constraint \eqref{eq:QCconstraint3}. To simplify the constraint, we focus on  $B_{12}$  and compute $B_{21}$  based on \eqref{eq:QCconstraint3}.  
Consequently, the COP problem can still be formulated as \eqref{eq:formulatedProblem}, where the parameter vector  $\vec{u}$  now includes the controller parameters  $A_K$,  $B_{K_2}$,  $B_{K_y}$, and  $C_K$, along with the direct coupling parameter $B_{12}$. The resulting controller is
\begin{equation} 
\begin{aligned} 
d\xi &= \begin{bmatrix} -9.81655374 & 0.47153487 \\ -1.02338419 & -8.90977693 \end{bmatrix} \xi dt \\ 
& \quad + \begin{bmatrix} -26.77010958 & -104.92638339 \\ 91.84663595 & -123.09708858 \end{bmatrix} dx(t) \\ 
& \quad + \begin{bmatrix} -0.00039029 & 0.00113423 \\ 0.00122208 & 0.00135909 \end{bmatrix} dw_{K_1} \\ 
& \quad + \begin{bmatrix} -3.00908255 & -2.67290672 \\ 0.25210814 & -5.69625236 \end{bmatrix} dw_{K_2} \\ 
& \quad + \begin{bmatrix} -0.65196678 & 0.66307307 \\ -1.26777189 & -0.10944131 \end{bmatrix} dy; \\ 
du(t) &= \begin{bmatrix} -0.00135909 & 0.00113423 \\ 0.00122208 & 0.00039029 \end{bmatrix} \xi dt + dw_{K_1},
\end{aligned} 
\end{equation}
with a performance parameter of  $J_\infty = 2.00646187$ and $k(\vec{u}) = 5.74329180 \times 10^{-18}$. It shows that the controller obtained by our algorithm performs better than the one proposed in \cite{zhang2010direct} ($J_\infty = 4.000049633093338$) when all constraints are satisfied.

\subsection{Direct Coupling, Indirect Coupling and Ideal Squeezers}
\label{sec:subModelC}
\noindent In the context of coherent quantum feedback control, ideal squeezers play a critical role in manipulating quantum noise by applying a Bogoliubov transformation to the input signals of a quantum optical system. The squeezing operation reduces uncertainty in one quadrature while increasing it in the conjugate quadrature, effectively enhancing the precision of quantum measurements. This operation is particularly beneficial in quantum optical systems where noise is a significant limitation.

Building upon the model presented in Section~\ref{sec:subModelB}, Zhang et al. \cite{zhang2012coherent} introduced ideal squeezers in the coherent feedback control of linear quantum optical systems. The appropriate combination of ideal squeezers and direct coupling enables the design of a closed-loop system that significantly enhances control performance. An ideal squeezing operation can be mathematically approximated through a Bogoliubov transformation, which defines the characteristics of an ideal squeezer. In the amplitude-phase quadrature representation, the ideal squeezer is depicted as a diagonal matrix:
\begin{equation}
    S = \begin{pmatrix}
    e^{-r} & 0 \\
    0 & e^{r}
    \end{pmatrix},
\end{equation}
where  $r$  is the squeezing parameter, $e^{-r}$  and  $e^{r}$  represent the squeezing and anti-squeezing effects on the quadrature components, respectively. This reflects that one quadrature is ``squeezed'' while the other is ``amplified''.



As illustrated in Fig.~\ref{ModelAll} (c), with the inclusion of squeezers, all input signals are squeezed before interacting with the system. The input signal $u$ to the plant is squeezed, resulting in the modified input $S_u u$. Similarly, the noise inputs to the controller, $w_{K_1}$ and $w_{K_2}$, are squeezed into $S_{w_{K_1}} w_{K_1}$ and $S_{w_{K_2}} w_{K_2}$, respectively. The input from the plant to the controller, $y$, is also squeezed, becoming $S_y y$. The dynamics of the quantum plant is described as 
\begin{equation}\label{eq:QCplantC}
    \begin{aligned}
    dx(t)   &= Ax(t)dt + B_{12}d\xi(t) + BS_{u}du(t) \\
            & \quad + B_w dw(t),  \qquad x(0) = x, \\ 
    dy(t)   &= Cx(t)dt + D_w dw(t),  \\
    z(t)    &= C_z x(t) + D_z S_{u}\beta_u(t).  
    \end{aligned}
\end{equation}
Correspondingly, the specific plant system is extended to 
\begin{alignat}{1}
    dx(t) &= \begin{bmatrix}0 & \triangle \\ -\triangle & 0 \end{bmatrix}xdt + B_{12}d\xi(t) + \begin{bmatrix}0 & 0 \\ 0 & -2\sqrt{k_1} \end{bmatrix}S_{u}du \nonumber
    \\ & \quad + \begin{bmatrix}0 & 0 & 0 & 0 \\ 0 & -2\sqrt{k_2} & 0 & -2\sqrt{k_3} \end{bmatrix} \begin{bmatrix}dw_1 \\ dw_2 \end{bmatrix}; \label{plant_e3} \\
    dy(t) &= \begin{bmatrix}2\sqrt{k_2} & 0 \\ 0 & 0 \end{bmatrix}xdt + dw_1.\nonumber
\end{alignat}
The controller to be designed is  
\begin{equation}
    \begin{aligned}
    d\xi(t) &= A_K \xi(t) dt + B_{21}dx(t) + B_{K_1} S_{w_{K_1}} dw_{K_1}(t) \\ 
    & \quad + B_{K_2} S_{w_{K_2}} dw_{K_2}(t) + B_{K_y} S_{y} dy(t), \\
    du(t) &= C_K \xi(t) dt + S_{w_{K_1}} dw_{K_1}(t).
    \end{aligned}
\label{controllerC}
\end{equation}

In the closed-loop system \eqref{eq:closeloop}, the matrices $\mathcal{A}$, $\mathcal{B}$, and $\mathcal{C}$ are modified as follows to account for the inclusion of direct coupling and ideal squeezers:
\begin{equation*}
\mathcal{A} = \begin{bmatrix} A & BS_{u}C_K + B_{12} \\ B_{K_y}S_{y}C + B_{21} & A_K \end{bmatrix};
\end{equation*}
\begin{equation*}
\mathcal{B} = \begin{bmatrix} B_w & BS_{u}S_{w_{K_1}} & 0_{2\times2} \\ B_{K_y}S_{y}D_w & B_{K_1}S_{w_{K_1}} & B_{K_2}S_{w_{K_2}} \end{bmatrix}; 
\end{equation*}
\begin{equation*}
\mathcal{C} = \begin{bmatrix} C_z & D_zS_{u}C_K \end{bmatrix}.
\end{equation*}

Zhang et al. \cite{zhang2012coherent} parameterized the controller and employed a two-step optimization algorithm to obtain a set of solutions, resulting in a well-performing controller with $J_\infty = 2.0004$.
In contrast, utilizing the proposed algorithm detailed in Section~\ref{sec:QCTDE}, we modify the algorithm parameters to better suit the increased complexity of the current problem. Specifically, the penalty step $p$ is set to $30$, and the scaling factor $\alpha$ is adjusted to $1,000$, while the remaining settings are consistent with Section~\ref{sec:subModelA}. This adjustment is necessary to accommodate the greater complexity of the system, requiring more computational effort.
We search for the controller parameters $A_K$, $B_{K_2}$, $B_{K_y}$, and $C_K$, direct coupling parameter $B_{12}$, as well as the squeezing parameters $S_u$, $S_{w_{K_1}}$, $S_{w_{K_2}}$, and $S_y$:
\begin{equation*}
A_{K} = 10^{2} \times \begin{bmatrix}-9.90333655 & -5.48587944 \\ 5.55069545 & -9.94910802 \end{bmatrix};
\end{equation*}
\begin{equation*}
B_{K_1} = 10^{-5} \times \begin{bmatrix}1.50413669 & -0.00057658 \\ 0.31866566 & -0.00000339 \end{bmatrix};
\end{equation*}
\begin{equation*}
B_{K_2} = \begin{bmatrix}-2.68060350 & -47.21025640 \\ 0.46462312 & -269.60700867 \end{bmatrix};
\end{equation*}
\begin{equation*}
B_{K_y} = \begin{bmatrix}-0.16862183 & -72.43115132 \\ 17.12554806 & -1.04866207 \end{bmatrix};
\end{equation*}
\begin{equation*}
C_K = 10^{-5} \times \begin{bmatrix}0.00000339 & -0.00057658 \\ 0.31866566 & -1.50413669 \end{bmatrix};
\end{equation*}
\begin{equation*}
B_{12} = 10^4 \times \begin{bmatrix}-1.44522297 & -0.91302597 \\ 0.92400557 & -1.43999043 \end{bmatrix};
\end{equation*}
\begin{equation*}
B_{21} = 10^4 \times \begin{bmatrix}1.43999043 & -0.91302597 \\ 0.92400557 & 1.44522297 \end{bmatrix};
\end{equation*}
\begin{equation*}
S_u = \begin{bmatrix}10.0 & 0.0 \\ 0.0 & 0.1 \end{bmatrix};
\quad 
S_y = \begin{bmatrix}2.06297180 & 0.0 \\ 0.0 & 0.48473760 \end{bmatrix};
\end{equation*}
\begin{equation*}
S_{w_{K_1}} = \begin{bmatrix}10.0 & 0.0 \\ 0.0 & 0.1 \end{bmatrix};  
S_{w_{K_2}} = \begin{bmatrix}9.99382305 & 0.0 \\ 0.0 & 0.10006181 \end{bmatrix}.
\end{equation*}
The controller shows excellent performance, achieving $J_\infty = 2.0000403964$ and $k(\vec{u}) = 2.0656859936 \times 10^{-17}$.

Based on this case, we conducts an ablation study to assess the impact of various configurations on the algorithm’s performance, as shown in Table.~\ref{tab:ablationStudy}. All methods utilize the Bet-and-Run Initialization from Section~\ref{sec:BRI} to obtain a promising initial solution. Row 0 represents the complete configuration, serving as the control group. Subsequent rows show adjustments with different settings for feasibility rules, penalty methods, search value scaling, and search range. Key observations from this study are as follows:
\begin{itemize}
    \item Strict Feasibility Rules (Row $1$): Replacing ``Relaxed Feasibility Rules'' with ``Strict Feasibility Rules'' (SFR) significantly reduces performance and increases runtime to four times that of the original algorithm. This result suggests that strict feasibility rules limit the algorithm’s optimization ability, making it harder to find satisfactory solutions.
    \item Fixed Penalty Factors (Rows $2-5$): Substituting ``Scheduled Dynamic Penalty'' with ``Fixed Penalty'' (FP) produces varied outcomes. Without penalty (FP($0$)), the performance index is best ($2.00003$), but the constraint is unsatisfied ($0.01$). When the penalty factor is 100 (FP($100$)), the configuration achieves relatively good performance and constraint satisfaction, although it still underperforms compared to the original algorithm. As the penalty factor gets larger, the performance gets worse. These results indicate that fixed penalty factors are less effective than dynamic penalties, as they lack the flexibility of the complete algorithm in balancing exploration and constraint satisfaction. Additionally, manually adjusting penalty factors can be challenging, as it is difficult to predict the optimal factor.
    \item No Search Value Scaling (Row $6$): When search values are not scaled, the search is restricted to a smaller range, leading to poor performance. 
    \item Fixed Search Range (Row $7$): Fixing the search range yields reasonably good results, but still underperforms compared to the complete algorithm due to the inflexibility of a fixed range.
    \item Search within the real fixed range (Rows $8-10$): Setting the search value scaling factor to $1$ and searching within the actual range reveal that performance declines as the range expands. Within the range of $[-10, 10]$ (Row $8$), both the performance index and the satisfaction with the constraints are well achieved. However, further expansion (Rows $9$ and $10$) results in poorer performance and, in some cases, no feasible solution. This observation underscores the importance of carefully selecting the search range. Although a smaller range may yield solutions, it restricts exploration of better solutions, whereas a larger range can hinder effective convergence to the optimal value.
\end{itemize}
\begin{table*}[htbp]
    \caption{Ablation study results showing the impact of configurations on algorithm performance.}
    \centering
    \begin{tabular}{c|c|c|c|c|c|c|c} \hline  
    \toprule
 & \textbf{Feasibility Rules}& \textbf{Penalty}& \textbf{Search Value Scaling}& \textbf{Search Range}& \textbf{Performance Index}& \textbf{Constraint}& \textbf{Time(s)}\\ \hline \hline  
         \rowcolor{Lavender}
         \textbf{0}&  {\textit{RFR}}&  {\textit{SDP}}&  {1,000}&  {Adaptive}&  {2.00004039640549}&  {2.07E-17}& 7947 \\ \hline  
         \textbf{1}&  \cellcolor{SkyBlue}\textit{SFR}&  \textit{SDP}&  1,000&  Adaptive&  7.69085396321596&  2.32E-16& 34068 \\ \hline  
         \textbf{2}&  \textit{RFR}&  \cellcolor{GreenYellow}\textit{FP}(0)&  1,000&  Adaptive&  2.00003041895946&  0.01& 7765\\ \hline  
         \textbf{3}&  \textit{RFR}&  \cellcolor{GreenYellow}\textit{FP}(100)&  1,000&  Adaptive&  2.00004069522938&  1.32E-16& 7630\\ \hline  
         \textbf{4}&  \textit{RFR}&  \cellcolor{GreenYellow}\textit{FP}(1E+05)&  1,000&  Adaptive&  2.00004123980377&  2.41E-16& 7970\\ \hline  
         \textbf{5}&  \textit{RFR}&  \cellcolor{GreenYellow}\textit{FP}(1E+10)&  1,000&  Adaptive&  2.26417536051186&  3.59E-17&10892\\ \hline  
         \textbf{6}&  \textit{RFR}&  \textit{SDP}&  \cellcolor{PineGreen}1&  Adaptive& 2.0074569809611 & 1.81E-16 & 8120\\ \hline  
         \textbf{7}&  \textit{RFR}&  \textit{SDP}&  1,000&  \cellcolor{Dandelion}[-10, 10]& 2.00004044351341 & 6.45E-16 &8233\\ \hline  
         \textbf{8}&  \textit{RFR}&  \textit{SDP}&  \cellcolor{YellowGreen}1&  \cellcolor{YellowGreen}[-10, 10]&  2.02024969423232&  1.39E-16& 8089\\ \hline  
         \textbf{9}&  \textit{RFR}&  \textit{SDP}&  \cellcolor{YellowGreen}1&  \cellcolor{YellowGreen}[-100, 100] &  54395.6251225345&  1.21E-14& 9044\\ \hline  
         \textbf{10}&  \textit{RFR}&  \textit{SDP}&  \cellcolor{YellowGreen}1&  \cellcolor{YellowGreen}[-1,000, 1,000]& N/A & N/A& 29821 \\ \hline \hline  
 \multicolumn{8}{c}{\textit{RFR}: Relaxed Feasibility Rules;  \quad \textit{SFR}: Strict Feasibility Rules; \quad   \textit{SDP}: Scheduled Dynamic Penalty; \quad \textit{FP}($\bullet$): Fixed Penalty factor $\bullet$}\\ 
    \bottomrule
    \end{tabular}
    \label{tab:ablationStudy}
\end{table*}

\section{conclusion}\label{sec:conclusion}
\noindent In this paper, we propose a generalized design approach for quantum controller design within the quantum coherent feedback control framework, leveraging evolutionary learning to address the unique challenges inherent in quantum systems. The proposed method integrates relaxed feasibility rules, a scheduled dynamic penalty function, adaptive range adjustment, and the “bet-and-run” initialization strategy. Through these enhancements, the algorithm effectively balances exploration and exploitation, and successfully identifies high-performance solutions while satisfying the feasibility constraints of quantum systems.

Numerical results across three different scenarios demonstrate that the proposed method outperforms traditional approaches in obtaining feasible quantum controllers. With minor adjustments to certain algorithm parameters, the method achieves exceptional performance tailored to diverse scenarios. Future work will focus on refining this approach for a broader range of quantum control systems and potentially extending its applicability to other domains, such as multi-objective quantum optimization problems.

\appendix[Differential Evolution]
\label{DEDetails}
\noindent DE iteratively optimizes solutions through mutation, crossover, and selection cycles. The implementation details below focus on our specific configuration. For comprehensive details about the canonical DE algorithm and theoretical foundations, readers may refer to \cite{storn1997differential,storn1996usage,price2006differential,das2010differential}.

\begin{itemize}
    \item \textit{Initialization}: 
    The population $\textbf{U}_0 = \{\vec{u}_{0,i}\}_{i=1}^{N_p}$ ($N_p=20$) is uniformly randomized within bounds \eqref{eq:QCfixedBounds}, ensuring initial search space coverage.
    
    \item \textit{Mutation}: 
    For each \textit{target vector} $\vec{u}_{g,i}$ (current candidate solution), generate a \textit{donor vector} $\vec{v}_{g,i}$ (mutated candidate) via:
    \begin{equation}
        \vec{v}_{g,i} = \vec{u}_{g,a} + F( \vec{u}_{g,b} - \vec{u}_{g,c}),
    \end{equation}
    where $\vec{u}_{g,a},\vec{u}_{g,b},\vec{u}_{g,c}$ are distinct population vectors. Scaling factor $F=0.85$ balances exploration-exploitation \cite{das2010differential}.
    
    \item \textit{Crossover}: 
    Blend donor $\vec{v}_{g,i}$ and target $\vec{u}_{g,i}$ to construct a \textit{trial vector} $\vec{z}_{g,i}$ (new candidate solution). This combines exploration from the donor vector and exploitation of the target vector:
    \begin{equation}
        z_{g,i}(j) = 
        \begin{cases} 
            v_{g,i}(j) & \text{if } r_j \leq \mathrm{CR} \text{ or } j = j_{\mathrm{rand}}, \\
            u_{g,i}(j) & \text{otherwise},
        \end{cases}
    \end{equation}
    where $r_j \in [0,1]$ is a uniform random number, $\mathrm{CR}=0.95$ controls donor component inheritance probability, and $j_{\mathrm{rand}}$ ensures at least one component from $\vec{v}_{g,i}$ to prevent stagnation.
    
    \item \textit{Selection}: 
    Choose the fitter vector between the trial vector and the target vector to ensure non-decreasing fitness:
    \begin{equation}
        \vec{u}_{g+1,i} = 
        \begin{cases}  
            \vec{z}_{g,i} & \text{if } f(\vec{z}_{g,i}) \leq f(\vec{u}_{g,i}), \\ 
            \vec{u}_{g,i} & \text{otherwise},
        \end{cases}
    \end{equation}
    where $f(\cdot)$ is the fitness function defined in \eqref{fitness3}, which evaluates the solution quality.
\end{itemize}

\bibliography{IEEEabrv,ref}
\bibliographystyle{IEEEtran}

\end{document}